\documentclass[a4paper, 11pt]{article}   
\usepackage{geometry}                		
\usepackage{graphicx}		
\usepackage{subcaption}		
\usepackage{amssymb}
\usepackage{amsmath}
\usepackage{mathtools}
\usepackage{algorithm}
\usepackage{authblk}
\usepackage{blindtext}
\usepackage{multirow}
\usepackage{algpseudocode}
\usepackage[colorlinks=true, linkcolor=blue, citecolor = blue]{hyperref}
\usepackage{booktabs}
\usepackage{tikz}
\usepackage[toc,title]{appendix}
\usetikzlibrary{patterns, arrows,positioning}
\newcommand{\bx}{\mathbf{x}}
\newcommand{\Cov}{\mathbb{C}\text{ov}}
\newcommand{\Cor}{\mathbb{C}\text{orr}}

\newcommand{\Exp}{\mathbb{E}}
\newcommand{\erf}{\text{erf}}

\title{Emulating computer models with step-discontinuous outputs using Gaussian processes}
\author[1, 2]{Hossein Mohammadi \thanks{Corresponding Author: h.mohammadi@exeter.ac.uk}}
\author[1, 2]{Peter Challenor}
\author[1, 2]{Marc Goodfellow}
\author[1]{Daniel Williamson}
\affil[1]{College of Engineering, Mathematics and Physical Sciences, University of Exeter, Exeter, UK}
\affil[2]{EPSRC Centre for Predictive Modelling in Healthcare, University of Exeter, Exeter, UK}
\date{}							
\begin{document}
	\maketitle
\abstract{In many real-world applications we are interested in approximating costly functions that are analytically unknown, e.g. complex computer codes. An emulator provides a ``fast" approximation of such functions relying on a limited number of evaluations. Gaussian processes (GPs) are commonplace emulators due to their statistical properties such as the ability to estimate their own uncertainty. GPs are essentially developed to fit smooth, continuous functions. However, the assumptions of continuity and smoothness is unwarranted in many situations. For example, in computer models where \emph{bifurcations} or \emph{tipping points} occur, the outputs can be discontinuous. 
	This work examines the capacity of GPs for emulating step-discontinuous functions. Several approaches are proposed for this purpose. Two ``special" covariance functions/kernels are adapted with the ability to model discontinuities. They are the neural network and Gibbs kernels whose properties are demonstrated using several examples. Another approach, which is called \emph{warping}, is to transform the input space into a new space where a GP with a standard kernel, such as the Mat\'ern family, is able to predict the function well. The transformation is performed by a parametric map whose parameters are estimated by maximum likelihood. The results show that the proposed approaches have superior performance to GPs with standard kernels in capturing sharp jumps in the ``true" function.}\\
{\bf Keywords:} Covariance kernel, Discontinuity, Emulator, Gaussian processes, Warping.
\section{Introduction}
\label{intro}
Computer models (or simulators) are widely used in many applications ranging from modelling the ocean and atmosphere \cite{adcroft2004, challenor2004} to healthcare \cite{birrell2011, proctor2013}. By simulating real-world phenomena, computer models allow us to better understand/analyse them as a complement to conducting physical experiments. However, on the one hand, the simulators are often ``black box" since they are available as commercial packages and we do not have access to their internal procedures. On the other hand, they are computationally expensive due to the fact that each simulation outcome is actually the solution of some complex mathematical equations, such as partial differential equations. 

One of the main purposes of using a computer model is to perform prediction. However, the accuracy of the prediction is questionable because simulators are simplifications of physical phenomena. In addition, due to factors such as lack of knowledge or measurement error, the inputs to the model are subject to uncertainty which yield uncertain outputs. Under this condition, decision makers need to know how good the prediction is. In other words, they need an estimation of the uncertainty propagated through the model \cite{oakley2004_1}. This entails running the simulator very many times which is impractical in the context of time-consuming simulators. To overcome this computational complexity, one can replace the simulator with an \emph{emulator} which is fast to run. 

Emulation is a statistical approach for representing unknown functions by approximating the input/output relationship based on evaluations at a finite set of points.
Gaussian process (GP) models (also known as kriging) are widely used to predict the outputs of a complex model and are regarded as an important class of emulators \cite{ohagan2006}. GPs are nonparametric probabilistic models that provide not only a mean predictor but also a quantification of the associated uncertainty. They have become a standard tool for the design and analysis of computer experiments over the last two decades. This includes uncertainty propagation \cite{oakley2004_2, lockwood2012}, model calibration \cite{kennedy2001, higdon2008}, design of experiments \cite{sacks1989, pronzato2012}, optimisation \cite{jones1998, brochu2010} and sensitivity analysis \cite{oakley2004_1, iooss2015}. 

GPs can be applied to fit any smooth, continuous function \cite{neal1998}. The basic assumption when using a GP emulator is that the unknown function depends smoothly on its input parameters. However, there are many situations where the model outputs are not continuous. It is very common in computer models that at some regions of the input space, a minor change in the input parameters leads to a sharp jump in the output. For example, models described by nonlinear differential equations often exhibit different modes (phases). Shifting from one mode to another relies on a different set of equations which raises a discontinuity in the model output. 

To our knowledge, there are only a few studies that investigate the applicability of GPs in modelling discontinuities. The reason may be due to the fact that they are essentially developed to model smooth and continuous surface forms. However, a natural way of emulating discontinuous functions is to partition the input space by finding discontinuities and then fit separate GP models within each partition. In \cite{caiado2015}, for example, a simulator with tipping point behaviour is emulated such that the boundary of the regions with discontinuity is found first and the simulator output is emulated separately in each region.  It is reported that finding the discontinuous regions is a time-consuming operation. 

The treed Gaussian process (TGP)  \cite{gramacy2008} is a popular model introduced by Gramacy and Lee. The TGP makes binary splits (parallel to the input axes) on the value of a single variable recursively such that each partition (leaf of the tree) is a subregion of the previous section. Then, an independent stationary GP emulator is applied within each section. The disadvantage of the TGP is that it requires many simulation runs which is not affordable in the context of computationally expensive simulators. A similar approach is presented in \cite{pope2018} where Voronoi tessellation is applied to partition the input space. The procedure uses the reversible jump Markov chain Monte Carlo \cite{green95} that is time-consuming. In \cite{ghosh2018} a two-step method is proposed for emulating cardiac electrophysiology models with discontinuous outputs. First a GP classifier is employed to detect boundaries of discontinuities and then the GP emulator is built subject to these boundaries. 

Here we provide an alternative perspective in which a single kernel is used to capture discontinuities. The advantage is that there is no need to detect discontinuous boundaries separately which is burdensome. The proposed methods include two nonstationary covariance functions, namely the neural network (NN) \cite{williams1997, raissi2018} and Gibbs kernels \cite{gibbs1997, paciorek2004}, and the idea of warping the input space \cite{calandra2016}. The NN kernel was first derived by Williams \cite{williams1997} and relies on the correspondence between GPs and single-layer neural networks with infinite number of hidden units (neurons) and random weight parameters \cite{neal1998}. As a result, the NN kernel is more expressive than standard kernels in modelling complex data structures. In the Gibbs kernel the parameter that regulates the correlation between observations is a function of the inputs which makes that kernel more flexible than the classical covariance functions. The warping technique has been already proven to be successful in modelling nonstationary functions, see e.g. \cite{sampson1992, snoek2014, marmin2018}. A warped kernel is obtained by applying a deterministic non-linear transformation to its inputs. This can be regarded as a special case of ``Deep Gaussian processes" which is a functional composition of multiple GPs \cite{damianou2013, salimbeni2017}.    
In this work we show how these techniques coming from machine learning can be employed in the field of computer experiments to emulate models that present very steep variations. 
\section{Overview of Gaussian process emulators}
\label{GP_emulators}
The random (or stochastic) process $Z = \left(Z(\bx)\right)_{\bx \in \mathcal{D}}$, i.e. a collection of random variables indexed by the set $\mathcal{D}$,  is a Gaussian process if and only if 
$
\forall \, N \in \mathbb{N}, ~ \forall \, \bx^j \in \mathcal{D}, ~ \left(Z(\bx^1), \dots, Z(\bx^N) \right)^\top
$ has a multivariate normal distribution on $\mathbb{R}^N$ \cite{GPML}. Let $\left(\Omega, \mathcal{B}, \mathbb{P} \right)$, where $\Omega$  is a sample space, $\mathcal{B}$ is a sigma-algebra and $\mathbb{P}$ is a probability measure, be the probability space on which $Z(\bx)$ is defined:
\begin{equation*}
	Z: (\bx, \omega)  \mapsto Z(\bx, \omega)~, ~ (\bx, \omega) \in \mathcal{D} \times \left(\Omega, \mathcal{B}, \mathbb{P} \right) .
\end{equation*} 
For a given $\omega_o \in \Omega$, $Z(\cdot, \omega_o)$ is called a \emph{sample path} (or \emph{realisation}) and for a given $\bx_o \in \mathcal{D}$, $Z(\bx_o, \cdot)$ is a Gaussian random variable.
In this framework, GPs can be regarded as the probability distribution over functions such that the function being approximated is considered as a particular realisation of the distribution. Herein, $f : \mathcal{D}  \mapsto \mathcal{F}$ denotes the unknown function that maps the input space $\mathcal{D} \subset \mathbb{R}^d$ to the output space $\mathcal{F}$. In this work, $\mathcal{F} = \mathbb{R}$.

A GP is fully determined by its mean function $\mu(\cdot)$ and covariance kernel $k(\cdot,\cdot)$ which are defined as:
\begin{table}[H]
	\centering
	\begin{tabular}{l l} 
		\multirow{2}{*}{$Z \sim \mathcal{GP}\left(\mu(\cdot), k(\cdot, \cdot)\right)$~;} & $\mu : \mathcal{D} \mapsto \mathbb{R}~,~ \mu(\bx) = \Exp\left[Z(\bx)\right] $\\ 
		& $k : \mathcal{D} \times \mathcal{D} \mapsto \mathbb{R}~, ~  k(\bx, \bx^\prime) = \Cov \left(Z(\bx), Z(\bx^\prime)\right) $.
	\end{tabular}
\end{table}
\noindent While $\mu$ could be any function, $k$ needs to be symmetric positive semidefinite. The function $\mu$ captures the global trend and $k$ controls the structure of sample paths such as differentiability, symmetry, periodicity, etc.  In this work, the GP mean is assumed to be an unknown constant which is estimated from data, see Equation (\ref{mu_estim}). The notation $\mu$ is (slightly abusively) used to denote the value of this constant.

Generally, covariance functions are divided into two groups: \emph{stationary} and \emph{nonstationary}. Stationary kernels depend only on the separation vector $\bx - \bx^\prime$. As a result, they are translation invariant in the input space:
\begin{equation}
	k(\bx, \bx^\prime) = k(\bx + \boldsymbol{\tau}, \bx^\prime + \boldsymbol{\tau}) \, ,~ \boldsymbol{\tau} \in \mathbb{R}^d.
\end{equation}
One of the most common covariance functions is the squared exponential (SE) kernel whose (separable) form is given by
\begin{equation}
	k_{SE}(\bx, \bx^\prime) = \sigma^2\prod_{i = 1}^{d}\exp \left( -\frac{\vert x_i - x^\prime_i\vert^2}{2l_i^2} \right) . 
\end{equation}
Here, the parameters $\sigma^2$ and $l_i$ are called \emph{process variance} and correlation \emph{length-scale} along the $i$-th coordinate, respectively. The former determines the scale of the amplitude of sample paths and the latter regulates how quickly the spatial correlation decays. In this paper, these parameters are estimated via maximum likelihood (ML) \cite{GPML, jones1998}, see Appendix \ref{sec_kernel}. Figure \ref{Fig:SE_kernel} shows the shape of the SE kernel and two sample paths with different length-scales. Another important class of stationary kernels is the Mat\'ern covariance function \cite{GPML}.
Nonstationary kernels are applied to model functions that do not have uniform smoothness within the input space and change significantly in some regions compared to others \cite{xiong2007}. In Sections \ref{sec_NNkernel} and \ref{sec_Gibskernel} two nonstationary covariance functions, namely the neural network and Gibbs kernels, are studied. 
\begin{figure}[htpb] 
	\centering
	\includegraphics[width=0.49\textwidth]{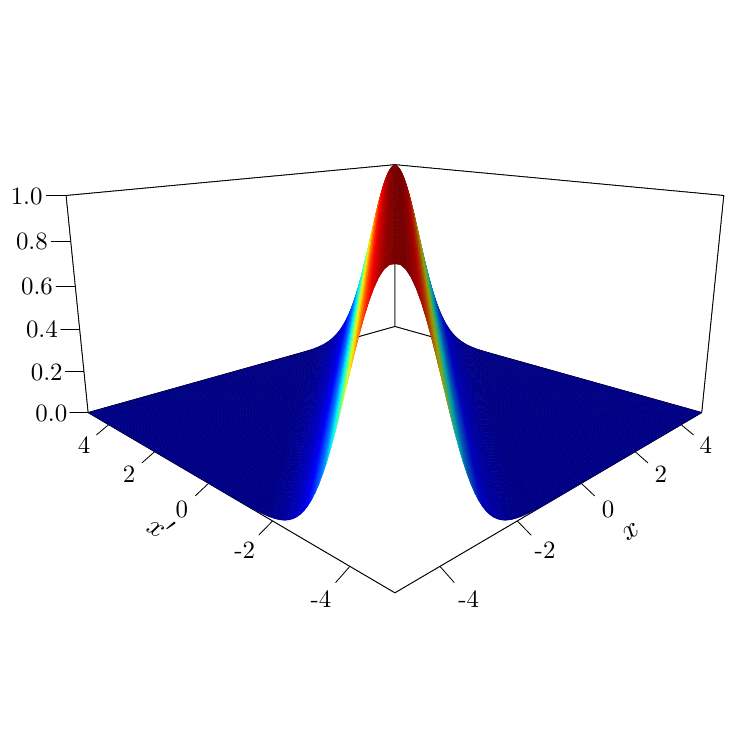} 
	\includegraphics[width=0.49\textwidth]{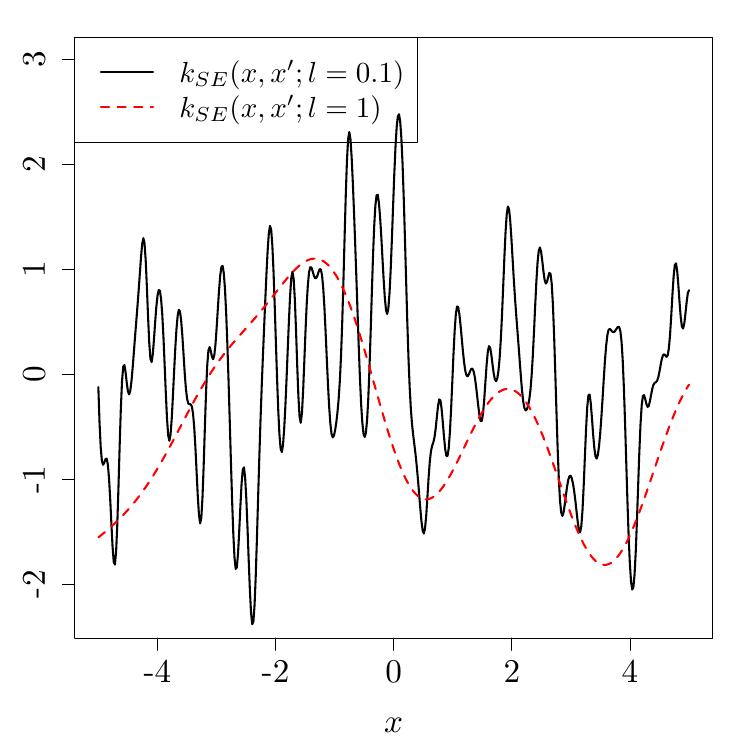}
	\caption{Left: shape of the squared exponential kernel. Right: sample paths corresponding to the SE kernel with $l= 0.1$ (solid) and $l = 1$ (red dashed). In both cases $\sigma^2 = 1$} 
	\label{Fig:SE_kernel}
\end{figure}

The GP prediction of $f$ is obtained by conditioning $Z$ on function evaluations. Let $\mathbf{X} = \left(\bx^1, \dots, \bx^n \right)^\top$ denote $n$ sample locations in the input space and $\mathbf{y} = \left(f(\bx^1), \dots, f(\bx^n) \right)^\top$ represent the corresponding outputs (observations). Together, the set $\mathcal{A} = \{\mathbf{X}, \mathbf{y}\}$ is called the \emph{training} set. The conditional distribution of $Z$ on $\mathcal{A}$ is again a GP 
\begin{equation}
	Z\vert \mathcal{A} \sim \mathcal{GP}\left(m(\cdot), c(\cdot, \cdot)\right) , 
\end{equation}
specified by
\begin{align}
	\label{post_mean}
	m(\bx) & = \hat{\mu} + \mathbf{k}(\bx)^\top  \mathbf{K}^{-1} \left(\mathbf{y} - \hat{\mu} \mathbf{1} \right)  \\
	c(\bx, \bx^\prime) & =  k(\bx, \bx^\prime) - \mathbf{k}(\bx)^\top  \mathbf{K}^{-1} \mathbf{k} (\bx^\prime) + \frac{\left(1 - \mathbf{1}^\top  \mathbf{K}^{-1} \mathbf{k} (\bx^\prime) \right)^2}{\left(\mathbf{1}^\top  \mathbf{K}^{-1} \mathbf{1} \right)}  .
	\label{post_var}
\end{align} 
Here, $\hat{\mu}$ is the ML estimate of $\mu$ obtained by \cite{GPML}
\begin{equation}
	\hat{\mu} = \frac{\mathbf{1}^\top \mathbf{K}^{-1} \mathbf{y}}{\mathbf{1}^\top \mathbf{K}^{-1} \mathbf{1}} .
	\label{mu_estim}
\end{equation}
Also, $\mathbf{k}(\bx) = \left(k(\bx, \bx^1), \dots, k(\bx, \bx^n)\right)^\top$, $\mathbf{K}$ is an $n \times n$ covariance matrix whose elements are: $\mathbf{K}_{i j} = k(\bx^i, \bx^j)~,~ \forall i, j~;~ 1  \leq i, j \leq n$ and $\mathbf{1}$ is a $n \times 1$ vector of ones.  We call $m(\bx)$ and $s^2(\bx) = c(\bx, \bx)$ the GP mean and variance which reflect the prediction and the associated uncertainty at $\bx$, respectively.

It can be shown that in the classic covariance functions such as Mat\'ern kernel where $k(\bx, \bx ; \sigma^2 = 1) = 1$, the predictive mean expressed by Equation (\ref{post_mean}) interpolates the points in the training set. Also, the prediction uncertainty (Equation (\ref{post_var})) vanishes there. To clarify, we obtain the prediction and the associated uncertainty at $\bx = \bx^j$, the $j$-th training point. In this case, $\mathbf{k}(\bx^j)$ is equivalent to the $j$-th column of the covariance matrix $\mathbf{K}$. Because $\mathbf{K}$ is a positive definite matrix, the term $ \mathbf{k}(\bx^j)^\top  \mathbf{K}^{-1}$ yields vector $\mathbf{e}_j =  \left(0, \dots, 0,  1, 0, \dots, 0 \right)$ whose elements are zero except the $j$-th element which is one. As a result  
\begin{align}
	\label{post_mean_interpolate}
	m(\bx^j) & = \hat{\mu}  +  \overbrace{\mathbf{k}(\bx^j)^\top \mathbf{K}^{-1}}^{\mathbf{e}_j }  (\mathbf{y} - \hat{\mu} \mathbf{1} ) = f(\bx^j) , \\
	\nonumber s^2(\bx^j) & =  k(\bx^j, \bx^j) - \mathbf{k}(\bx^j)^\top  \mathbf{K}^{-1} \mathbf{k} (\bx^j) \\
	& + \frac{\left(1 - \mathbf{1}^\top  \mathbf{K}^{-1} \mathbf{k} (\bx^j) \right)^2}{\mathbf{1}^\top  \mathbf{K}^{-1} \mathbf{1}} = 0 ,
	\label{post_var_interpolate}
\end{align} 
since $\mathbf{k} (\bx^j) = (k(\bx^j, \bx^1), \dots, \overbrace{k(\bx^j, \bx^j)}^{\sigma^2}, \dots, k(\bx^j, \bx^n) )^\top$. 
\section{Neural network kernel}
\label{sec_NNkernel}
In this section we first show how the NN kernel is derived from a single-layer neural network with infinite number of hidden units. Let $\tilde{f}(\bx) $ be a neural network with $N_h$ units that maps inputs to outputs according to
\begin{equation}
	\tilde{f}(\bx) = b + \sum_{j = 1}^{N_h} v_j h(\bx ; \mathbf{u}^j) , 
	\label{neural_net}
\end{equation} 
where $b$ is the intercept, $v_j$s are weights to the units, $h(\cdot)$ represents the transfer (activation) function in which $\mathbf{u}^j$ represents the weight assigned to the input $\bx$.
Suppose $b$ and every $v_j$ have zero mean Gaussian distribution with variances $\sigma^2_b$ and $\sigma^2_v / N_h$, respectively. If $\mathbf{u}^j$s have independent and identical distribution, then the mean and covariance of $\tilde{f}(\bx)$ are ($\mathbf{w}$ represents all weights together, i.e. $\mathbf{w} = \lbrack b, v_1, \ldots v_{N_h}, \mathbf{u}^1, \ldots,   \mathbf{u}^{N_h}\rbrack$)
\begin{align}
	\Exp_{\mathbf{w}} \big[\tilde{f}(\bx) \big] &= 0 \, , \\
	\nonumber \Cov \left(\tilde{f}(\bx), \tilde{f}(\bx^\prime) \right) &= \Exp_{\mathbf{w}} \big[ (\tilde{f}(\bx) - 0) ( \tilde{f}(\bx^\prime) - 0 ) \big] \\
	\nonumber &= \sigma^2_b + \frac{1}{N_h} \sum_{j = 1}^{N_h} \sigma^2_v \Exp_{\mathbf{u}} \big[h(\bx; \mathbf{u}^j) h(\bx^\prime; \mathbf{u}^j) \big]   \\
	&= \sigma^2_b + \sigma^2_v \Exp_{\mathbf{u}} \big[h(\bx; \mathbf{u})  h(\bx^\prime; \mathbf{u}) \big] .
	\label{neural_net_kernel}
\end{align}
Since $\tilde{f}(\bx)$ is the sum of independent random variables, it tends towards a normal distribution as $N_h \to \infty$ according to the central limit theorem. In this situation, any collection $\big\{\tilde{f}(\bx^1), \ldots , \tilde{f}(\bx^N)| ~\forall N \in \mathbb{N} \big\}$ has a joint normal distribution and $\tilde{f}(\bx)$  becomes a zero mean Gaussian process with a covariance function specified in Equation (\ref{neural_net_kernel}). 

\begin{figure}[htpb] 
	\centering
	\includegraphics[width=0.49\textwidth]{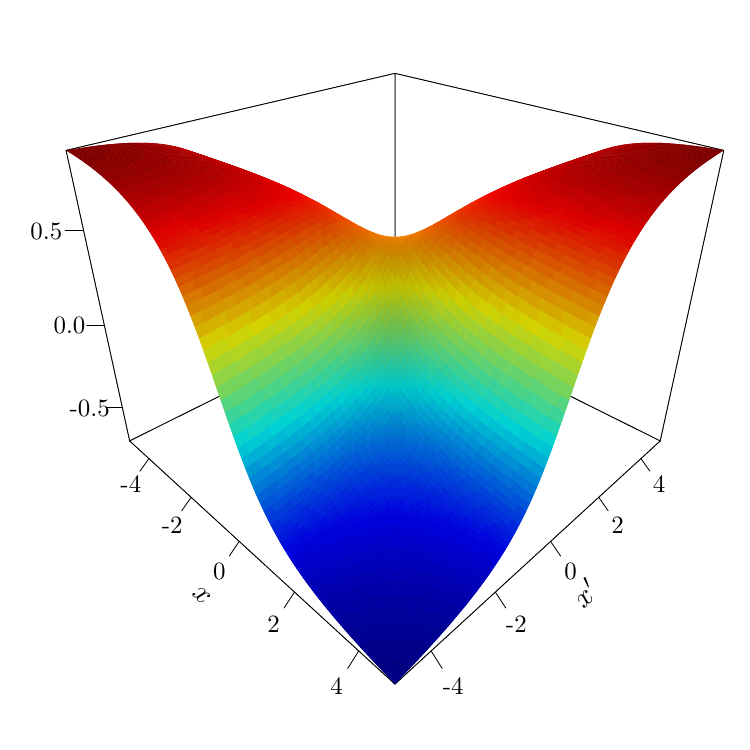}
	\includegraphics[width=0.49\textwidth]{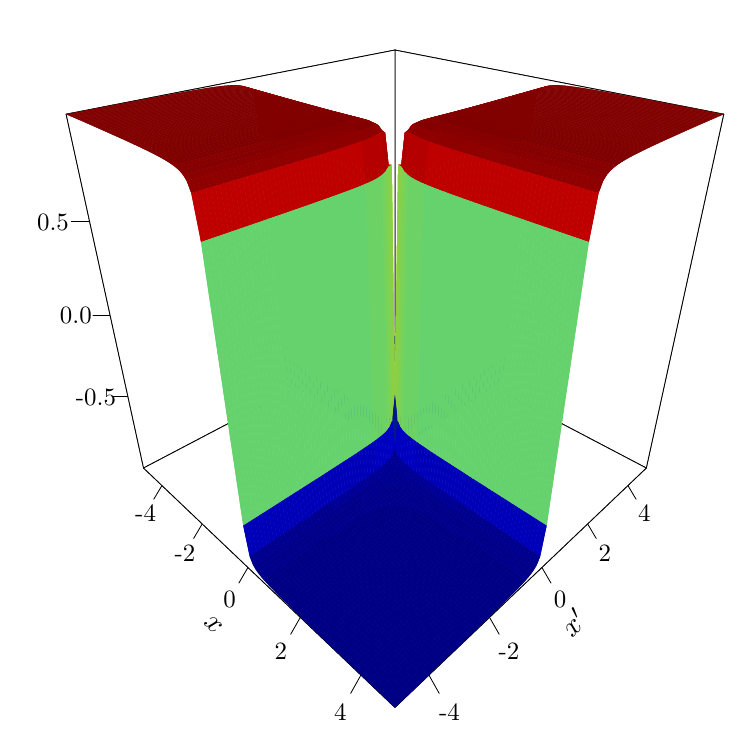} \\
	\includegraphics[width=0.49\textwidth]{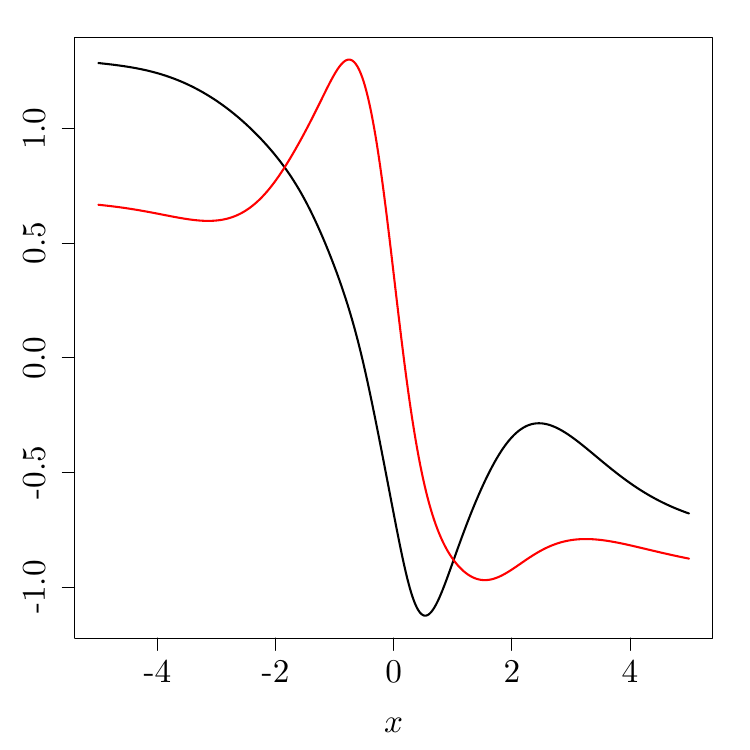}
	\includegraphics[width=0.49\textwidth]{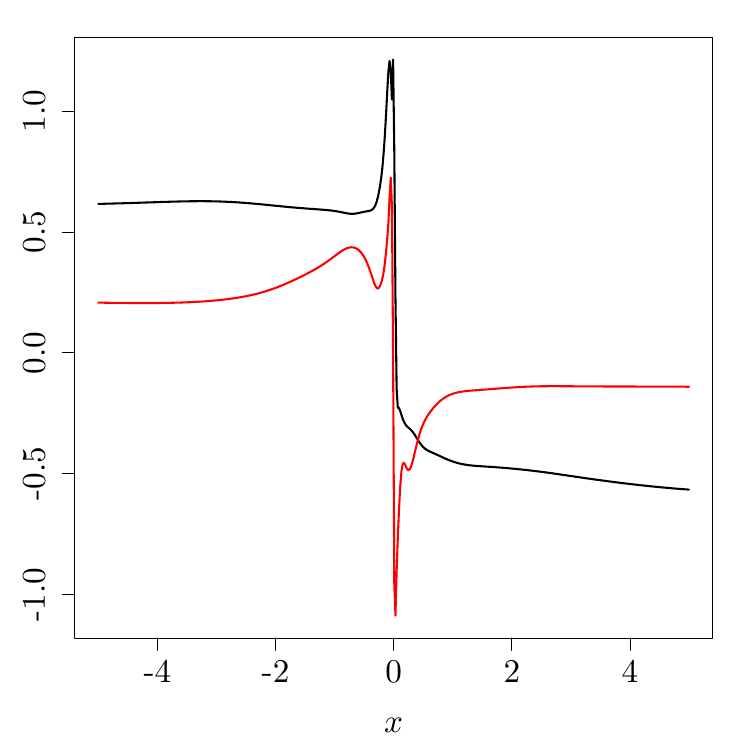}
	\caption{Top: The shapes of the NN kernel for two different values of $\sigma_1$: $1$ (left) and $50$ (right). Bottom: Two sample paths  corresponding to the kernels on top. The kernel with $\sigma_1 = 50$ is a more suitable choice for modelling discontinuities. Here, $\sigma = \sigma_0 = 1$.} 
	\label{NN_kernel_sample_path}
\end{figure}

The neural network kernel is a particular case of the covariance structure expressed by Equation (\ref{neural_net_kernel}) such that $h(\bx; \mathbf{u}) = \erf \left(u_0 + \sum_{i = 1}^d u_i x_i \right)$ \cite{williams1997}. Here, $\erf(\cdot)$ is the \emph{error function}: $\erf(x) = \frac{2}{\sqrt\pi} \int_{0}^{x} \exp(-t^2) dt$ and $\mathbf{u} \sim \mathcal{N}(\mathbf{0}, \boldsymbol{\Sigma})$ in which $\boldsymbol{\Sigma}$ is a diagonal matrix with elements $\sigma^2_0, \sigma^2_1, \dots, \sigma^2_d$ as the variances of $u_0, u_1, \dots, u_d$. This choice of the activation function leads to the neural network kernel given by 
\begin{equation}
	k_{NN}(\bx, \bx^\prime)  = \frac{2\sigma^2}{\pi}\arcsin\left( \frac{2\tilde{\bx}^\top\boldsymbol{\Sigma}\tilde{\bx}^\prime}{\sqrt{(1 + 2\tilde{\bx}^\top\boldsymbol{\Sigma}\tilde{\bx})(1 + 2\tilde{\bx}^{\prime\top}\boldsymbol{\Sigma}\tilde{\bx}^\prime)}} \right) .
	\label{NN_kernel}
\end{equation}
where $\tilde{\bx} = (1, x_1, \dots, x_d)^\top$ is the augmented input vector. The length-scale of the $i$-th coordinate is of order $1/ \sigma_i$; the larger $\sigma_i$, the sample functions vary more quickly in the $i$-th coordinate \cite{mackay1998, GPML}. This is illustrated in Figure \ref{NN_kernel_sample_path} where the shapes of the NN kernel for two different values of $\sigma_1$ and the corresponding sample paths are plotted. 
The NN kernel is nonstationary, see Figure \ref{NN_kernel_sample_path} and also Equation (\ref{NN_kernel}) which does not depend on $\bx - \bx^\prime$. It can take negative values contrary to classic covariance functions such as the SE kernel depicted in Figure \ref{Fig:SE_kernel}. In this kernel due to the superposition of the function $\erf(u_0 + u_1 x)$, sample paths tend to constant values for large positive or negative $x$ \cite{GPML}. Also, the correlation at zero distance is not one:
\begin{equation}
	\Cor(Z(\bx), Z(\bx)) = k_{NN}(\bx, \bx ; \sigma^2 = 1)= \frac{2}{\pi}\arcsin\left( \frac{2\tilde{\bx}^\top\boldsymbol{\Sigma}\tilde{\bx}}{1 + 2\tilde{\bx}^\top\boldsymbol{\Sigma}\tilde{\bx}} \right) < 1 ,
\end{equation}
since $\arcsin\left( \frac{2\tilde{\bx}^\top\boldsymbol{\Sigma}\tilde{\bx}}{1 + 2\tilde{\bx}^\top\boldsymbol{\Sigma}\tilde{\bx}} \right) < \pi/2$. Thus, the mean predictor obtained by the NN kernel does not interpolate the points in the training data and the prediction variances are greater than zero there.

\begin{figure}[htpb] 
	\centering
	\includegraphics[width=0.49\textwidth]{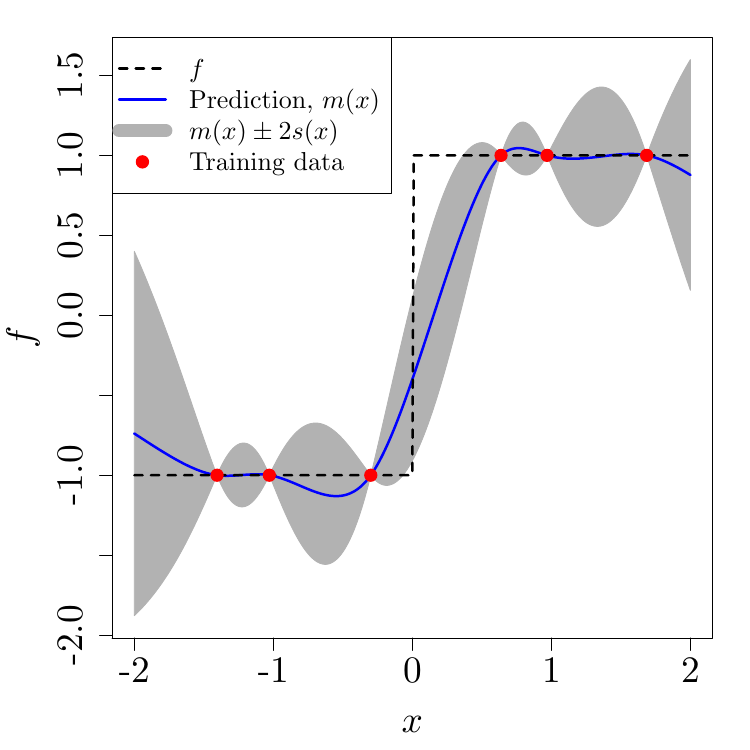}
	\includegraphics[width=0.49\textwidth]{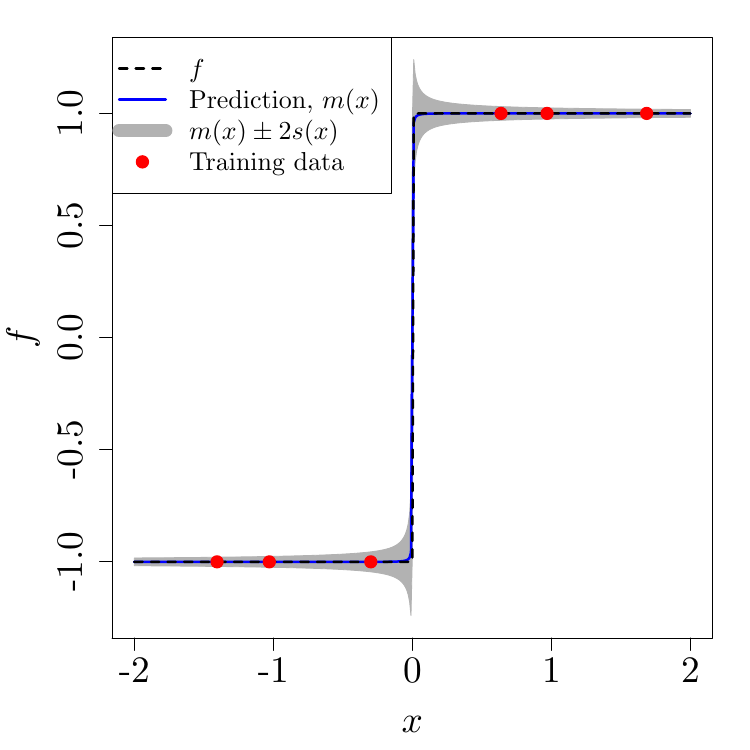} \\
	\includegraphics[width=0.49\textwidth]{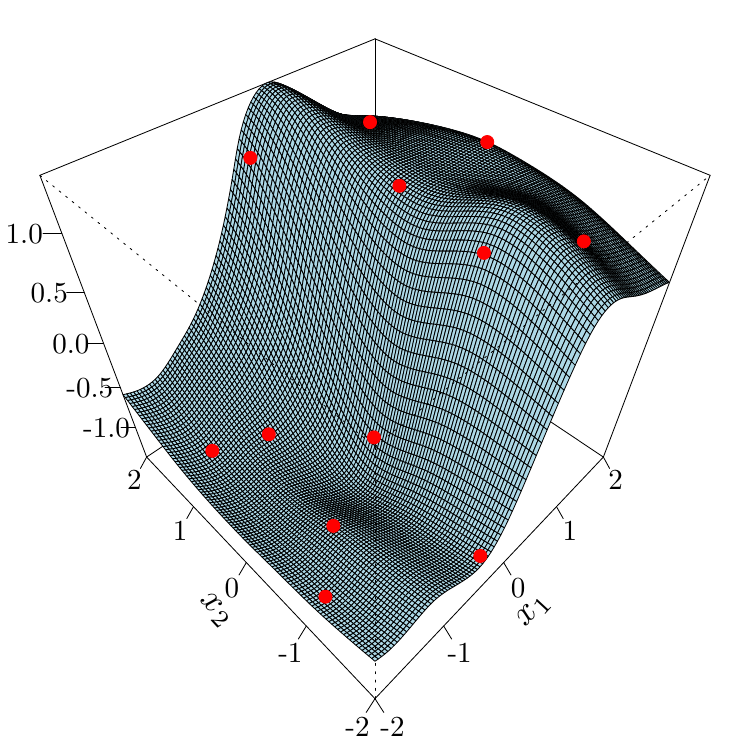}
	\includegraphics[width=0.49\textwidth]{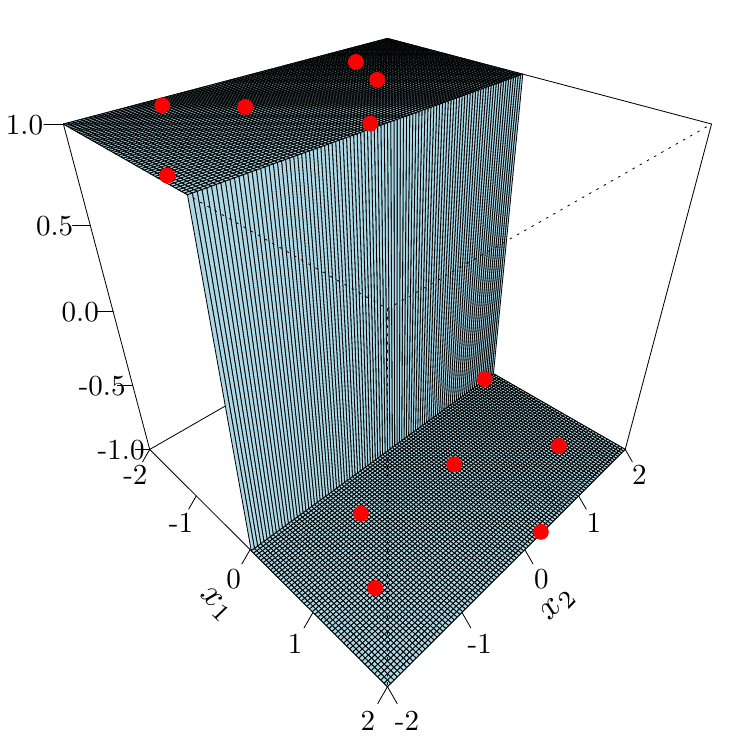}
	\caption{Emulation of the step-function $f$ defined in (\ref{step_fun}) with the Mat\'ern 3/2 (left panel) and NN (right panel) kernels in $1D$ (top row) and $2D$ (bottom row). Red points are the training data. In the NN kernel, $1D$: $~\hat{\sigma}_1 = 10^{3}$ which is the upper bound in the likelihood optimisation. $2D$: $\hat{\sigma}_{1}=10^{3}$ (corresponding to $x_1$) and $\hat{\sigma}_{1}=10^{-2}$ (corresponding to $x_2$) which is the upper bound in the likelihood optimisation.} 
	\label{step_fun_emul}
\end{figure}

Figure \ref{step_fun_emul} compares the Mat\'ern 3/2 and NN kernels in modelling a step-function defined as 
\begin{equation} 
	f(\bx) = 
	\begin{cases}
		-1 & x_1 \leq 0 \\
		1 &  x_1 >0~.
	\end{cases}
	\label{step_fun}
\end{equation}
As can be seen, the NN kernel has superior performance to Mat\'ern 3/2 in both $1D$  and $2D$ cases. The predictive mean of the GP with Mat\'ern 3/2  neither captures the discontinuity nor performs well in the flat regions. In the NN kernel, the ML estimation of the parameter that controls the horizontal scale of fluctuation, i.e. $\sigma_1$, takes its maximum possible value which is $10^{3}$.

\begin{figure}[htpb] 
	\centering
	\includegraphics[width=0.49\textwidth]{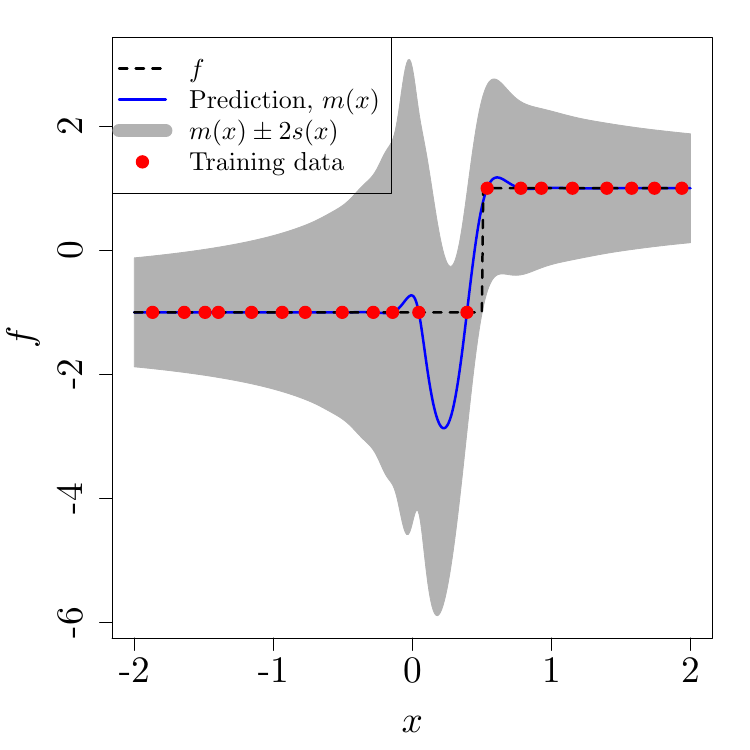}
	\includegraphics[width=0.49\textwidth]{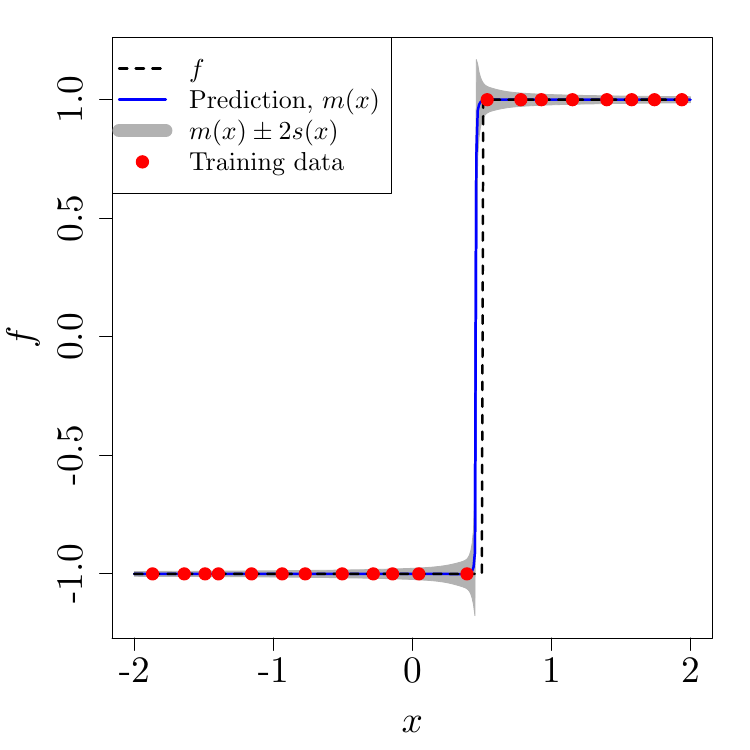} 
	\caption{Left: The NN kernel given by Equation (\ref{NN_kernel}) is not able to well-approximate $f$ (dashed red) whose discontinuity is at 0.5. Right: The modified NN kernel based on Equation (\ref{NN_kernel_perturb}) can well-model $f$. The ML estimation of $\tau$ is $0.457$ which is the estimated location of the discontinuity.} 
	\label{perturb_step_fun_emul}
\end{figure}

Figure \ref{perturb_step_fun_emul} illustrates a function whose step-discontinuity is located at $x = 0.5$. As can be seen from the picture on the left of Figure \ref{perturb_step_fun_emul}, the NN kernel is not able to model $f$ well. This problem can be solved if the NN kernel is modified as follows  
\begin{equation}
	k(x, x^\prime)  = \frac{2\sigma^2}{\pi}\arcsin\left( \frac{2\tilde{\bx}_{\tau}^\top\boldsymbol{\Sigma}^{-1}\tilde{\bx}_{\tau}^\prime}{\sqrt{(1 + 2\tilde{\bx}_{\tau}^\top\boldsymbol{\Sigma}^{-1}\tilde{\bx}_{\tau})(1 + 2\tilde{\bx}_{\tau}^{\prime\top}\boldsymbol{\Sigma}^{-1}\tilde{\bx}_{\tau}^\prime)}} \right),
	\label{NN_kernel_perturb}
\end{equation}
where $\tilde{\bx}_{\tau} = (1,  x - \tau)^\top$ and $\tau$ is estimated together with other parameters using ML. In this case, $\hat{\tau} = 0.457$ which is an estimation for the location of the discontinuity.
\section{Gibbs kernel}
	\label{sec_Gibskernel}
Mark Gibbs \cite{gibbs1997} in his PhD thesis derived the following covariance function: 
\begin{equation}
	k_{Gib}(\bx, \bx^\prime)  = \sigma^2 \prod_{i= 1}^{d} \left( \frac{2l_i(\bx)l_i(\bx^\prime) }{l_i(\bx)^2 + l_i(\bx^\prime)^2} \right)^{1/2} \exp \left(- \sum_{i = 1}^{d} \frac{(x_i - x_i^\prime)^2}{l_i(\bx)^2 + l_i(\bx^\prime)^2} \right),
\end{equation}
where $l_i(\cdot)$ is a length-scale function in the $i$-th input dimension. These length-scales can be any arbitrary positive functions of $\bx$. This allows the kernel to model sudden variations in the observations: a process with Gibbs kernel is smooth at regions of the input space where the length-scales are relatively high and it changes rapidly where the length-scales reduce. Note that the correlation is one when $\bx = \bx^\prime$, i.e.  $k_{Gib}(\bx, \bx) = 1$.
In this work, we use the same length-scale functions for all dimensions:
\begin{equation}
	k_{Gib}(\bx, \bx^\prime)  = \sigma^2 \left( \frac{2l(\bx)l(\bx^\prime)}{l^2(\bx) + l^2(\bx^\prime)} \right)^{d/2} \exp \left( -\frac{\sum_{i=1}^{d} (x_i - x_i^\prime)^2}{l^2(\bx) + l^2(\bx^\prime)}  \right).
\end{equation}

\begin{figure}[htpb] 
	\centering
	\includegraphics[width=0.3\textwidth]{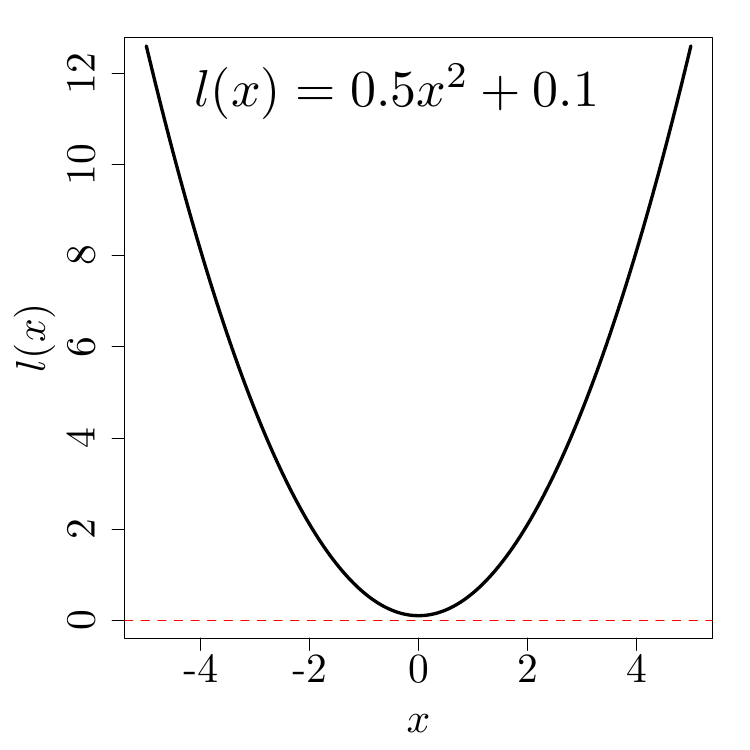}
	\includegraphics[width=0.34\textwidth]{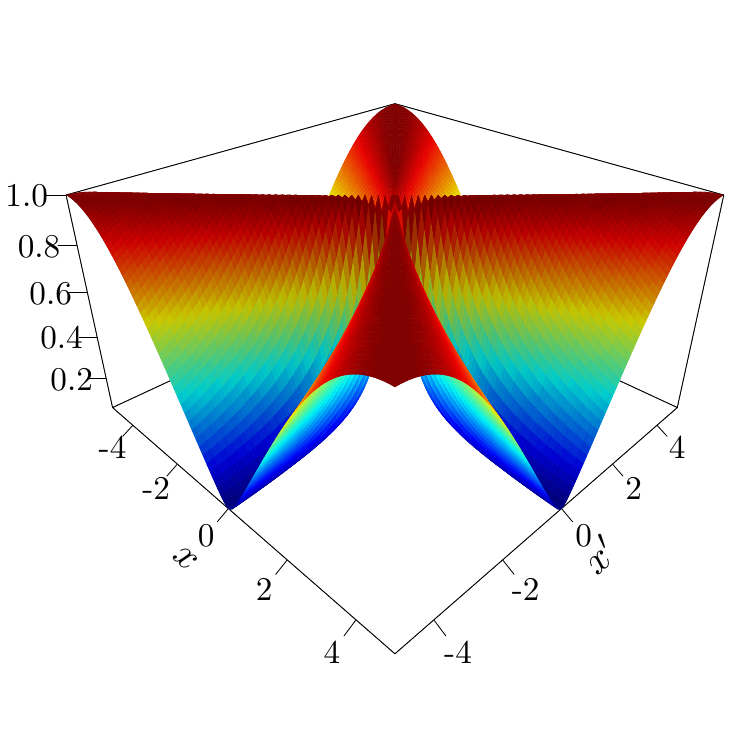} 
	\includegraphics[width=0.3\textwidth]{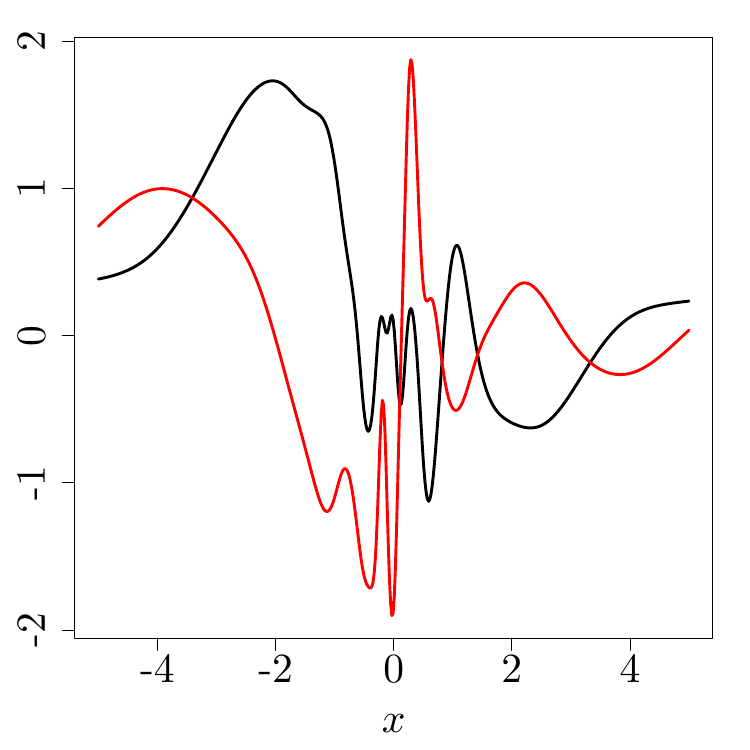} 
	\includegraphics[width=0.3\textwidth]{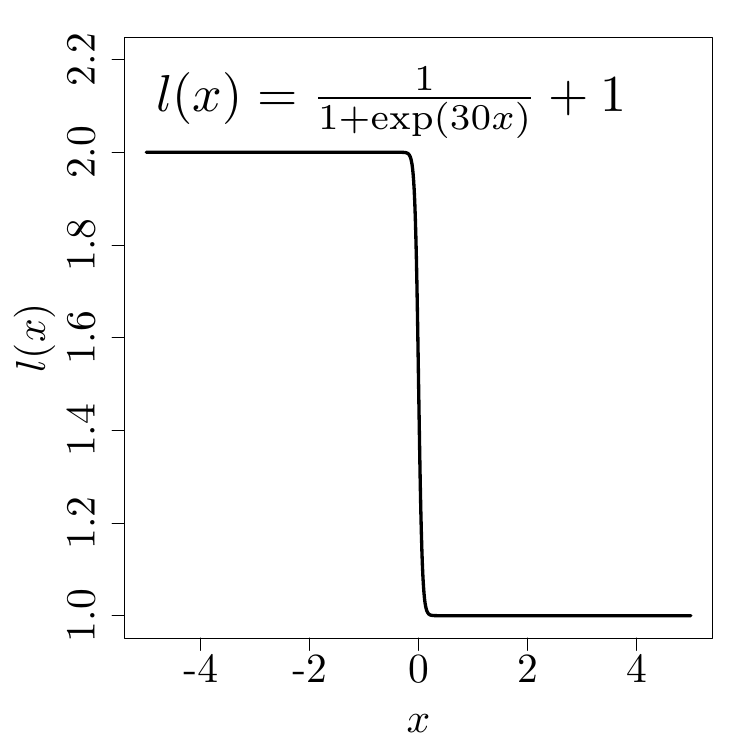}
	\includegraphics[width=0.34\textwidth]{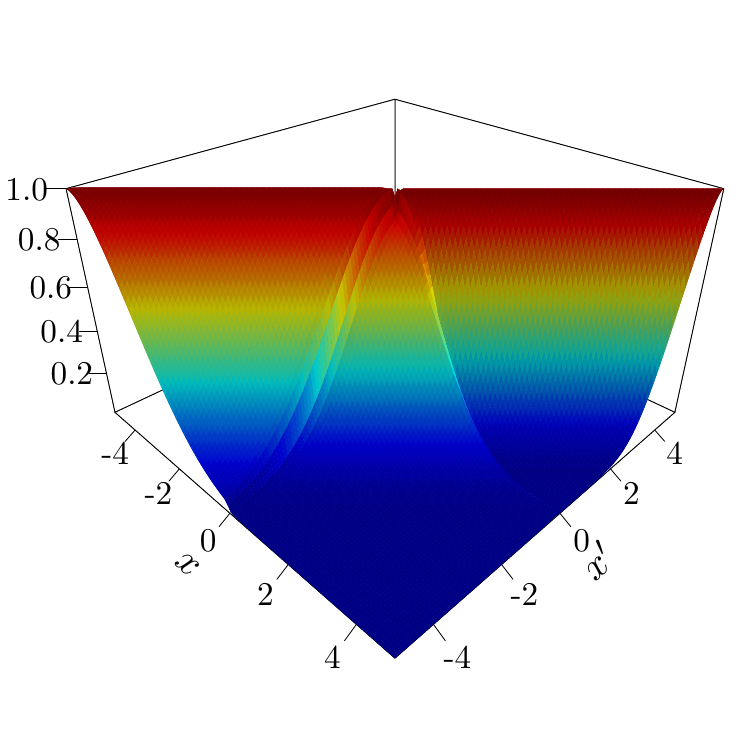} 
	\includegraphics[width=0.3\textwidth]{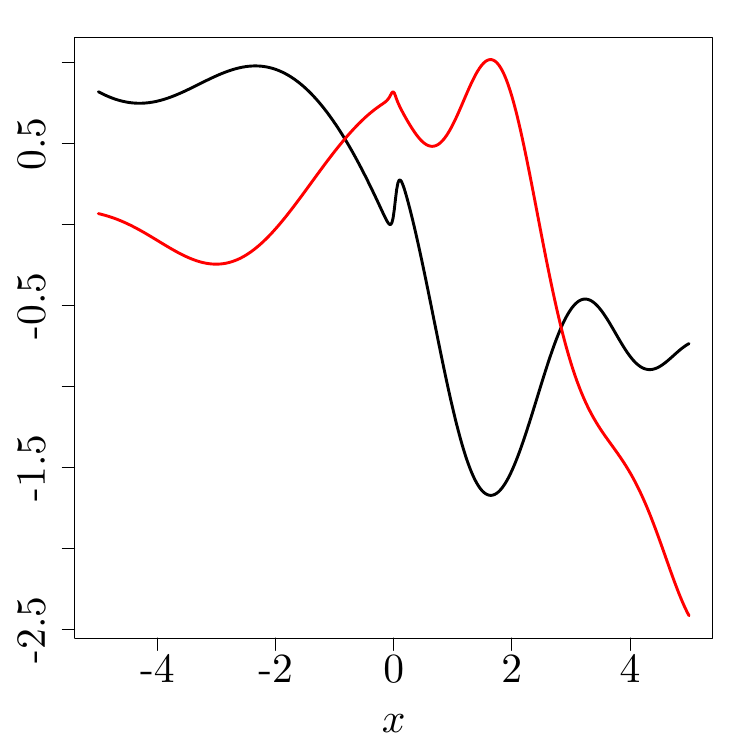} 
	\includegraphics[width=0.3\textwidth]{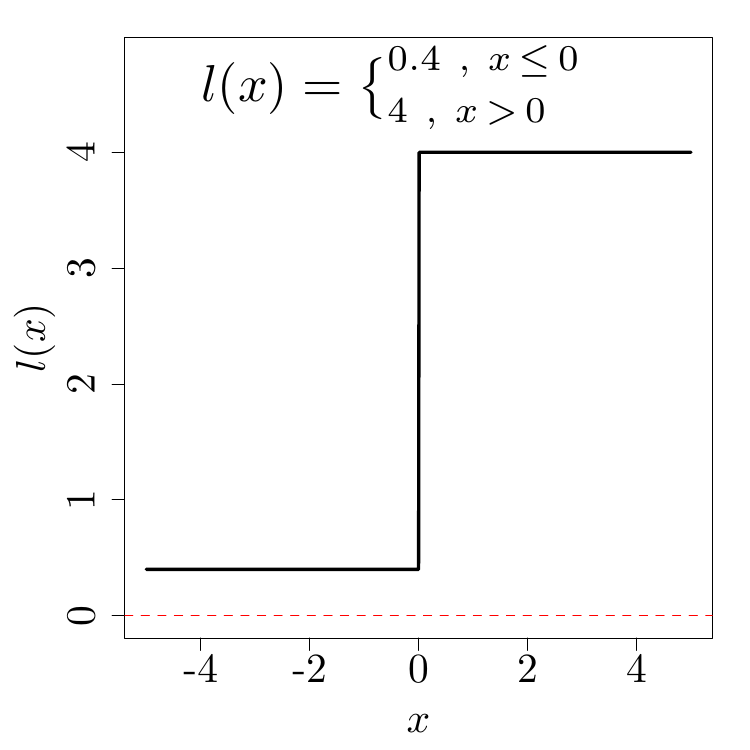}
	\includegraphics[width=0.34\textwidth]{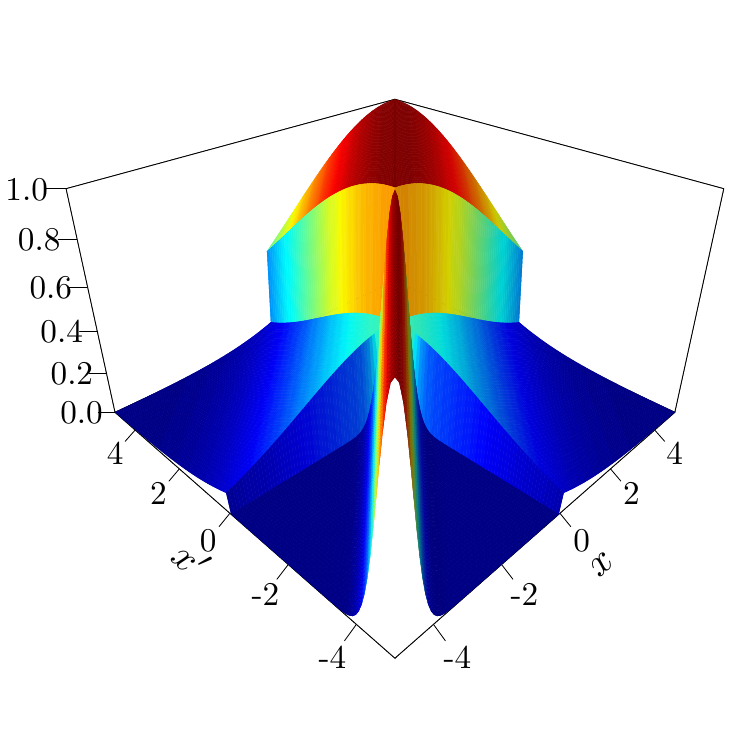} 
	\includegraphics[width=0.3\textwidth]{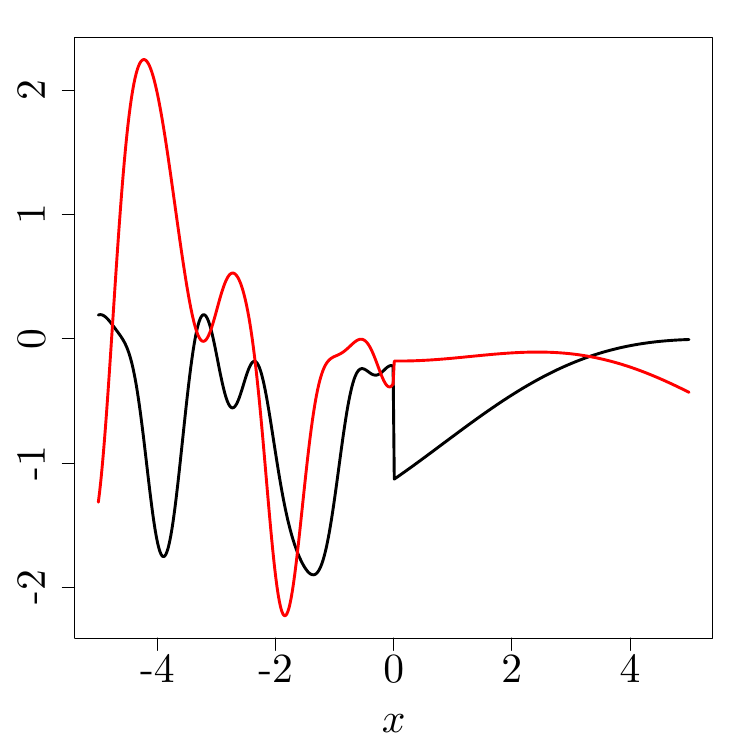} 
	\caption{Left panel: three different length-scale functions. Middle panel: shapes of the Gibbs kernel based on the corresponding length-scale functions. Right panel: two GP sample paths with the Gibbs kernel on the left. With the Gibbs kernel, one can model both nonstationary (first row) and discontinuous (second and third rows) functions.} 
	\label{gibbs_kernel}
\end{figure}

\begin{figure}[htpb] 
	\centering
	\includegraphics[width=0.49\textwidth]{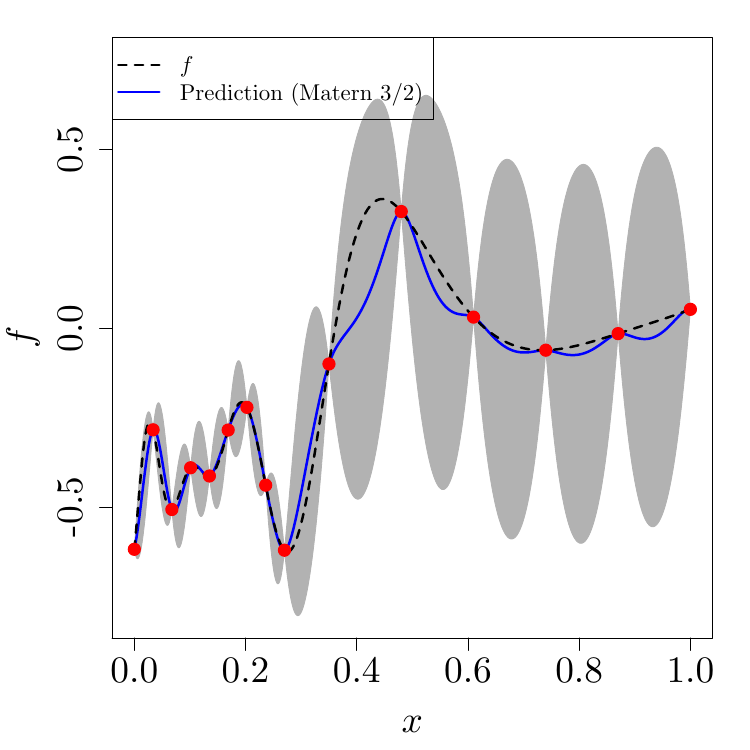}
	\includegraphics[width=0.49\textwidth]{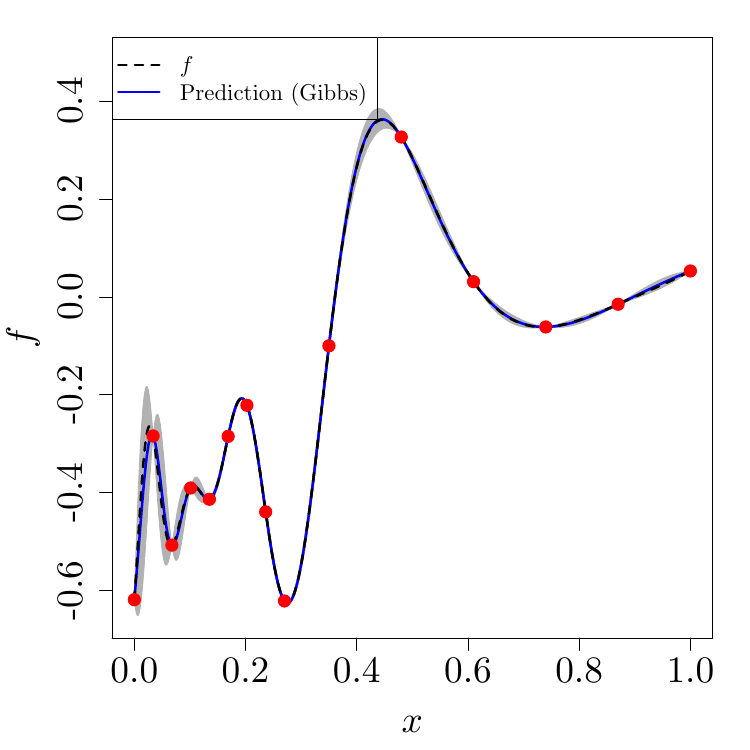} 
	\caption{GP prediction (solid blue) of a nonstationary function (dashed) with the Mat\'ern 3/2 (left) and Gibbs (right) kernels. The function is defined as $f(x) = \sin\left(30(x - 0.9)^4\right) \cos\left(2(x - 0.9)\right) + \frac{(x - 0.9)}{2}$ \cite{xiong2007} which varies more quickly in the region $x \in [0 , 0.3 ]$ than in the region $[0.3 , 1]$. The length-scale function used in the Gibbs kernel is $l(x) = c_1x^2 + c_2$ whose parameters are estimated by ML: $\hat{c_1} \approx 45.63$ and $\hat{c_2} \approx 0.11$. The shaded area represents the prediction uncertainty and the red points are the training data.} 
	\label{nonstation_gibbs}
\end{figure}

Figure \ref{gibbs_kernel} shows the shapes of the Gibbs kernel for three different length-scale functions and corresponding sample paths. As can be seen, it is possible to model both nonstationary and discontinuous functional forms with the Gibbs kernel if a suitable length-scale function is chosen. For example, the nonstationary function depicted in Figure \ref{nonstation_gibbs} varies more quickly in the region $x \in [0 , 0.3 ]$ than in the region $[0.3 , 1]$. Thus, a suitable length-scale function should have ``small" values when $x \in [0 , 0.3 ]$ and larger values when $x \in [0.3 , 1]$. The length-scale used for the Gibbs kernel (right picture) is of the form $l(x) = c_1x^2 + c_2$ whose unknown parameters $c_1$ and $c_2$ are estimated by ML. This choice of the length-scale allows the GP to predict $f$ with a higher accuracy in comparison to the GP with the Mat\'ern 3/2 kernel (left picture). The estimated parameters of the length-scale are $\hat{c}_1 \approx 45.63$ and $\hat{c}_2 \approx 0.11$ which are in line with the nonstationarity of $f$.

In order to model discontinuities, one can employ the Gibbs kernel with a sigmoid shaped length-scale. The length-scale functions we use in our experiments (see Section \ref{sec_result}) are all sigmoid functions, specifically:
\begin{itemize}
	\item[(i)] Error function: $\erf(c_1 \mathbf{e}_j \bx ) + c_2 ; ~ c_2 > 1$
	\item[(ii)] Logistic function: $\frac{1}{1 + \exp(c_1 \mathbf{e}_j \bx)} + c_2 ;~ c_2 > 0$
	\item[(iii)] Hyperbolic tangent: $\tanh(c_1 \mathbf{e}_j \bx ) + c_2 ; ~ c_2 > 1$
	\item[(iv)] Arctangent: $\arctan(c_1 \mathbf{e}_j \bx ) + c_2 ; ~ c_2 > \pi/2$
\end{itemize}   
which have all been modified slightly by adding a constant $c_2 > 0$ to make $l(\bx)$ strictly positive. The parameter $c_1$ controls the slope of the transition in the sigmoid function. Both $c_1$ and $c_2$ are estimated by ML. All components of the vector $\mathbf{e}_j$ are zero except the $j$-th one which is 1. This vector determines the $j$-th axis in which the function is discontinuous. 
\section{Transformation of the input space (warping)}
\label{sec_warping}
In this section, \emph{warping} or \emph{embedding} is studied as an alternative approach to emulate functions with discontinuities. The method first uses a non-linear parametric function to map the input space into a feature space. Then, a GP with a standard kernel is applied to approximate the map from the feature space to the output space \cite{mackay1998, calandra2016}.  A similar idea is used in \cite{snelson2004} where the transformation is performed on the output space to model non-Gaussian processes. 

\begin{figure}[htpb] 
	\centering
	\includegraphics[width=0.3\textwidth]{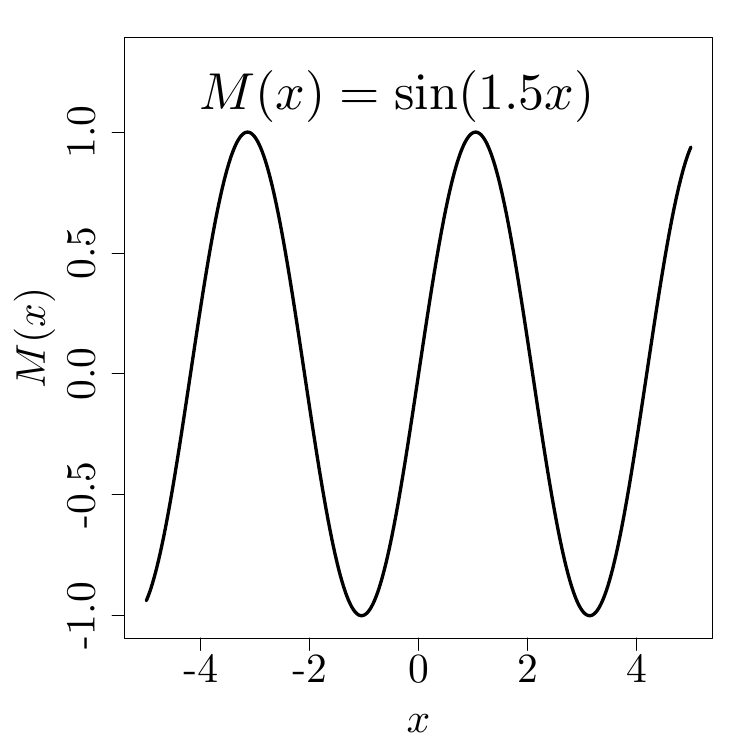}
	\includegraphics[width=0.34\textwidth]{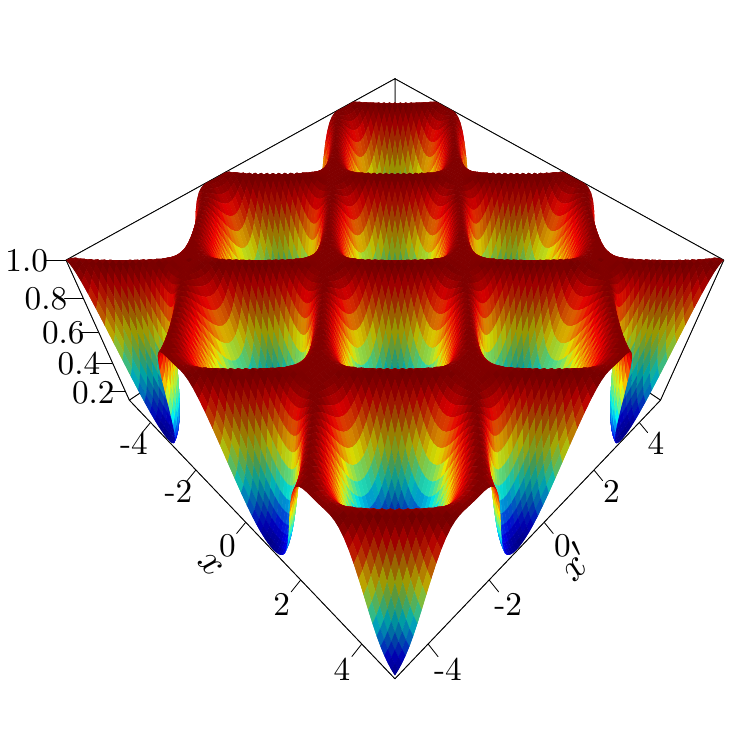} 
	\includegraphics[width=0.3\textwidth]{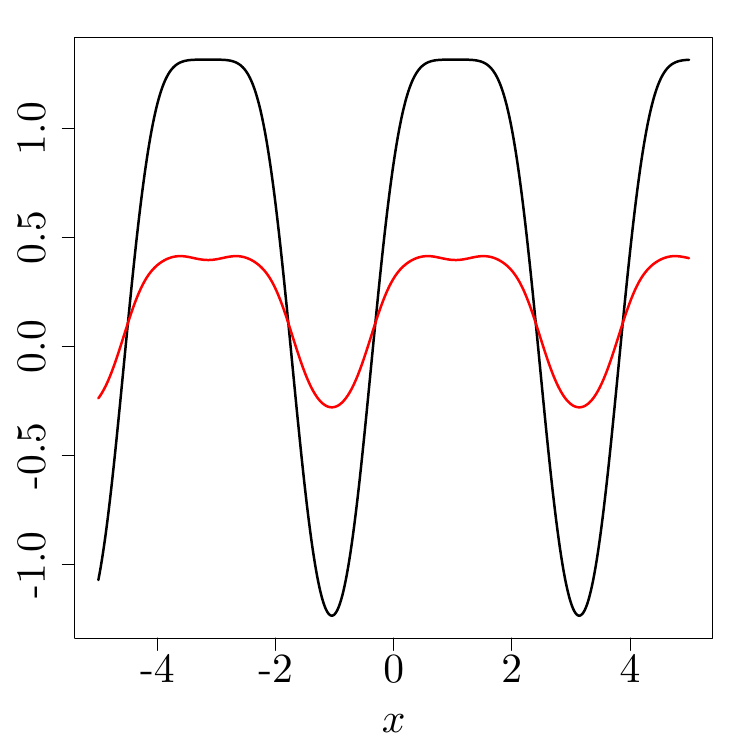} 
	\includegraphics[width=0.3\textwidth]{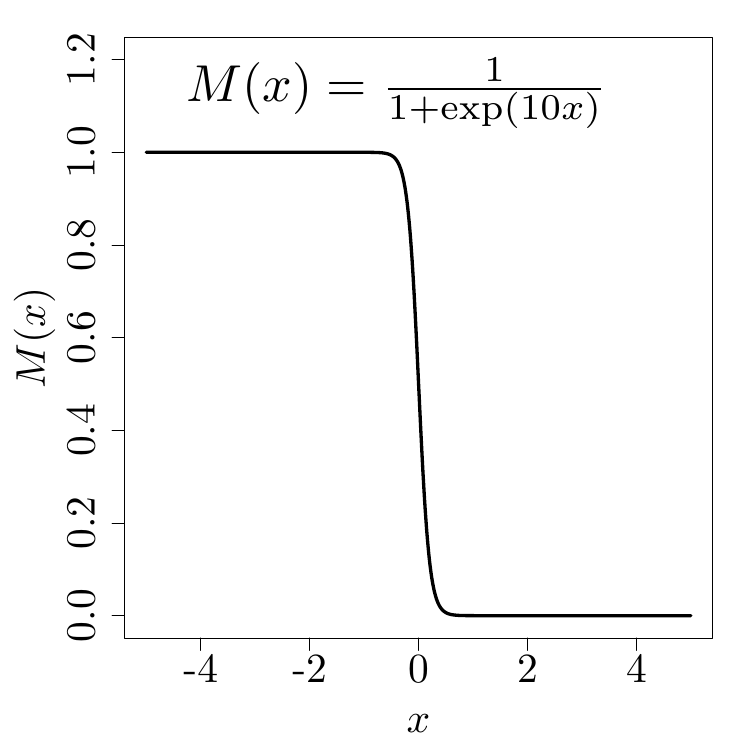}
	\includegraphics[width=0.34\textwidth]{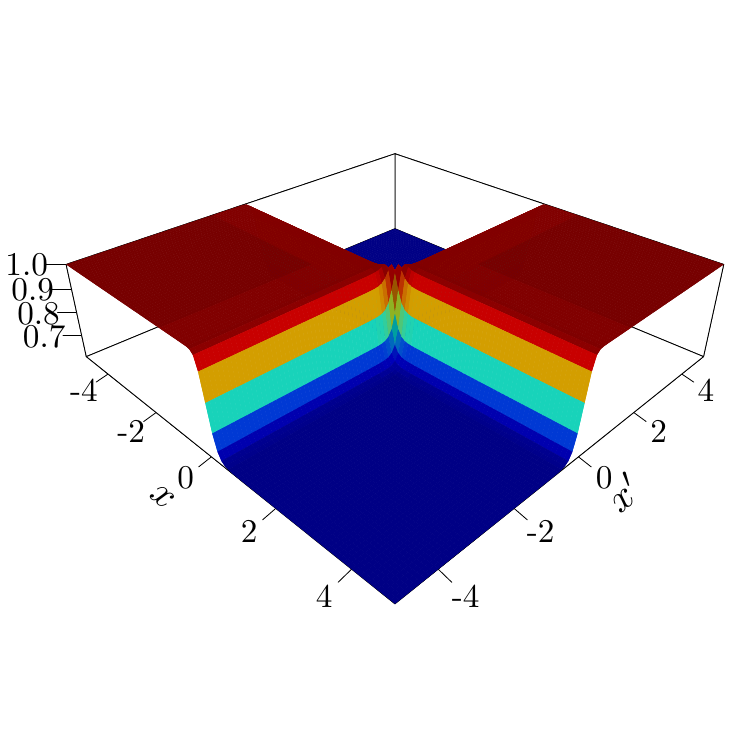} 
	\includegraphics[width=0.3\textwidth]{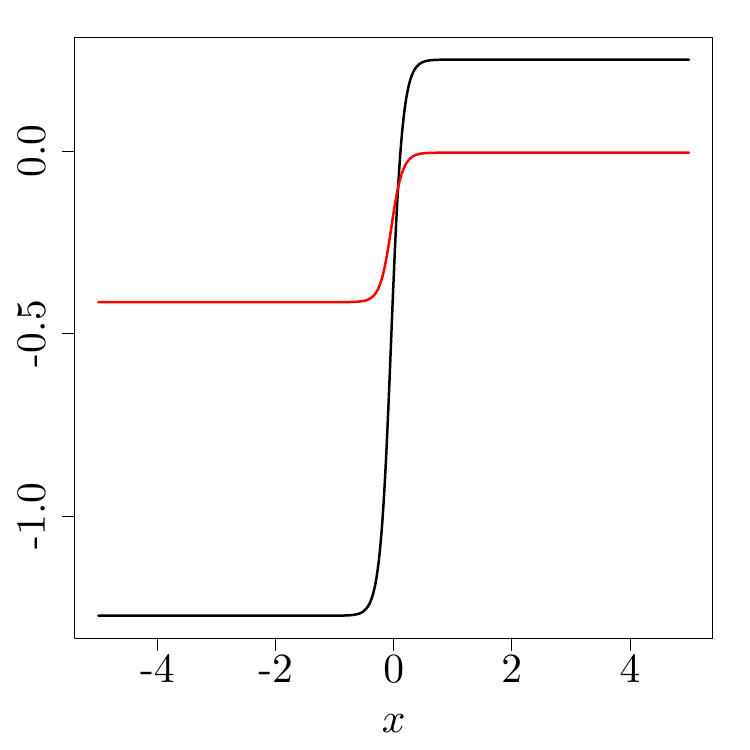} 
	\caption{Left panel: two different transformation functions, $M(x)$. Middle panel: shapes of the warped kernel $\tilde{k}\left(x, x^\prime\right) = k\left(M(x), M(x^\prime)\right)$ in which $k$ is the squared exponential kernel. Right panel: two sample paths of a GP with the covariance function $\tilde{k}$. As can be seen, a sigmoid transformation function is a suitable choice for modelling step-discontinuities.} 
	\label{fig_warp_kernel}
\end{figure}

In warping, we assume that $f$ is a composition of two functions
\begin{equation}
	f = G \circ M : ~ M : \mathcal{D} \mapsto \mathcal{D}^\prime ~, ~ G : \mathcal{D}^\prime \mapsto \mathcal{F} , 
\end{equation}
where $M$ is the transformation function and $\mathcal{D}^\prime$ represents the feature space. The function $G$ is approximated by a GP relying on the training set $\{\tilde{\mathbf{X}}, \mathbf{Y} \}$ in which $\tilde{\mathbf{X}} = \left(M(\bx^1), \ldots, M(\bx^n)\right)^\top$. Notice that $\mathcal{D}$ and $\mathcal{D}^\prime$ need not have the same dimensionality \cite{mackay1998}. For example, if the squared exponential kernel $k_{SE} :  \mathbb{R} \times \mathbb{R} \mapsto \mathbb{R}$ is composed with $M(x) = \left[\cos(\frac{2\pi x}{T}), \, \sin(\frac{2\pi x}{T})\right]^\top \in \mathbb{R}^2$, the result is a periodic kernel with period $T$ \cite{seeger2003, hajighasemi2014}. In practice, a parametric family of $M$ is selected and its parameters are estimated together with the kernel parameters via ML. Such modelling is equivalent to emulate $f$ with a GP whose covariance function $\tilde{k}$ is
\begin{equation}
	\tilde{k}(\bx, \bx^\prime) = k\left(M(\bx), M(\bx^\prime) \right) .
	\label{warp_kernel}
\end{equation}
Note that $\tilde{k}$ is generally nonstationary even if $k$ is a stationary kernel, see Figure \ref{fig_warp_kernel}. The prediction (conditional mean) and the associated uncertainty (conditional variance) of a warped GP are calculated in the same way as Equations (\ref{post_mean}) and (\ref{post_var}). 

According to Figure \ref{fig_warp_kernel} (second row), a sigmoid transformation is a suitable choice for modelling step-discontinuities. The sigmoid functions given in Section \ref{sec_Gibskernel} are used as the transformation mappings in our experiments in the next section. The unknown parameter of the map, i.e. $c_1$, is estimated by the ML method together with other kernel parameters such as the length-scales and process variance. 
\section{Numerical examples}
\label{sec_result}
In this section, the performance of the proposed methods in modelling step-discontinuities is compared with the standard kernels, i.e. Mat\'ern 3/2 and squared exponential. The step-function given by Equation (\ref{step_fun}) is used as the test function in dimensions 2 and 5. The input space is $\mathcal{D} = [-2, 2]^d$.  Four sigmoid functions are employed as the length-scales of the Gibbs kernel, $l(\bx)$, and the transformation maps, $M(\bx)$, in the warping method. The sigmoid functions are: error, logistic, hyperbolic tangent and arctangent whose analytical expressions are given in Section \ref{sec_Gibskernel}. The covariance kernel, $k$, in the warping approach is squared exponential. 

\begin{figure}[htpb] 
	\centering
	\includegraphics[width=0.49\textwidth]{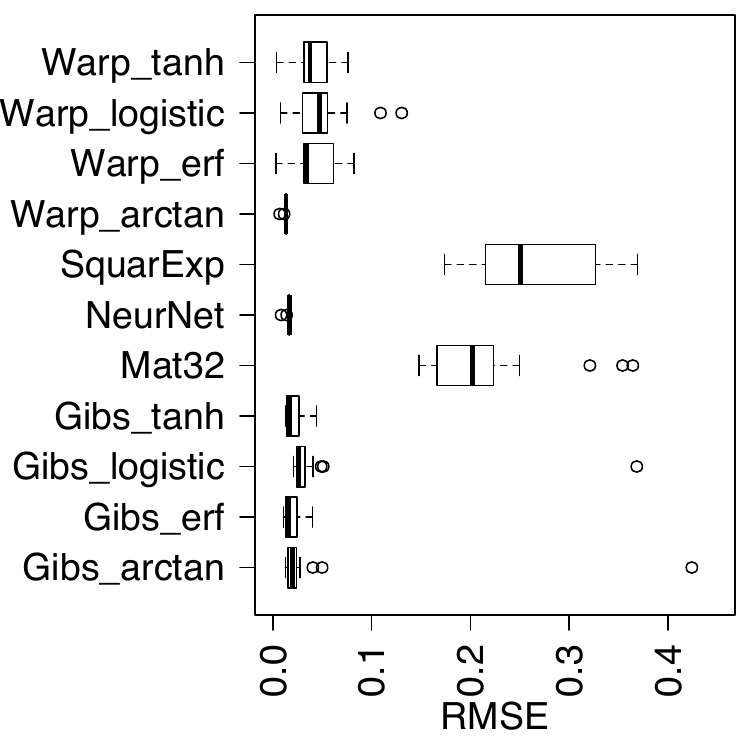}
	\includegraphics[width=0.49\textwidth]{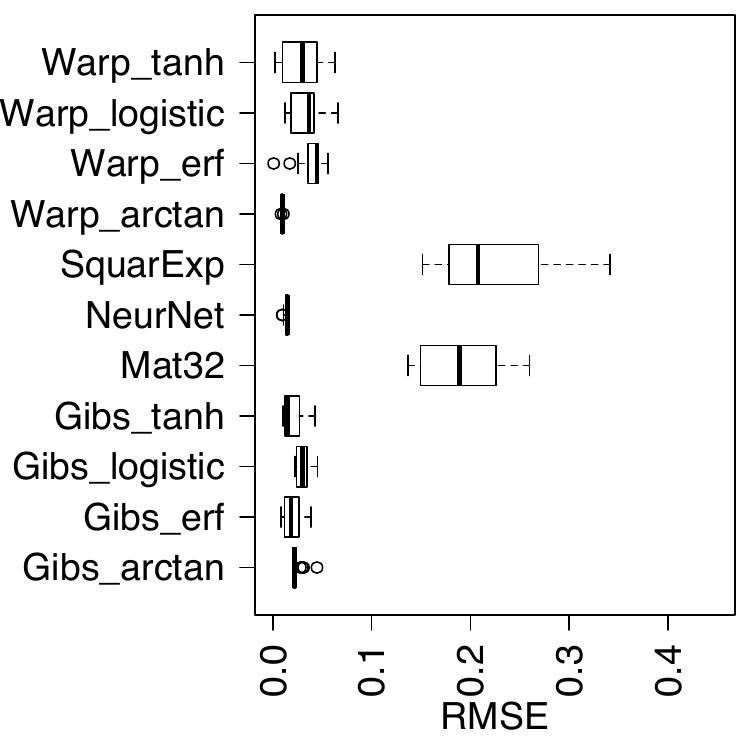} 
	\caption{Box-plot of RMSEs associated with the prediction of the step-function (Equation (\ref{step_fun})) in $2D$ (left) and $5D$ (right) using the methods presented in this work: neural network kernel, Gibbs kernel and warping. Four sigmoid functions (error, logistic, hyperbolic tangent and arctangent) are considered as the length-scale of the Gibbs kernel and the transformation function in the warping approach. The two standard kernels, i.e. squared exponential (SquarExp) and Mat\'ern 3/2 (Mat32), have the worst prediction performances. The best results are obtained by  the neural network kernel, Gibbs kernel (with arctangent) and warping (with arctangent).} 
	\label{fig_box_plot}
\end{figure}

The accuracy of the prediction is measured by the \emph{root mean square error (RMSE)} criterion defined as
\begin{equation}
RMSE = \sqrt{\frac{1}{n_t} \sum_{t = 1}^{n_t} \left(f(\bx_t) - \hat{f}(\bx_t) \right)^2} ,
\end{equation}
where $\bx_t$ and $n_t$ represent a test point and the size of test set, respectively. In our experiments $n_t = 1000$.
There are 20 different training sets and for each set, each method produces one prediction. All training sets are of size $10d$ and ``space-filling", meaning that the sample points are uniformly spread over the input space. They are obtained by the \texttt{maximinESE\_LHS} function implemented in the R package \emph{DiceDesign} \cite{DiceDesign}. The results are shown in Figure \ref{fig_box_plot}. 

As can be seen, the squared exponential (SquarExp) and Mat\'ern 3/2 (Mat32) kernels have the worst prediction performances. The neural network kernel (NeurNet) can model the step-function well and has one of the best RMSEs in our experiments. Generally, the GP model with the Gibbs kernel outperforms the warping technique. The arctangent function is a suitable choice as the length-scale of the Gibbs kernel and the transformation map in the warping approach. In both cases, the RMSE associated with the logistic function is (on average) the largest in comparison to other sigmoid functions. 
\section{Conclusions}
Gaussian processes are mainly used to predict smooth, continuous functions. However, there are many situations in which the assumptions of continuity and smoothness do not hold. In computer experiments, it is common that the output of a complex computer code has discontinuity, e.g. when bifurcations or tipping points occur. This paper deals with the problem of emulating step-discontinuous functions using GPs. Several methods, including two covariance kernels and the idea of transforming the input space (warping), are proposed to tackle this problem. The two covariance functions are the neural network and Gibbs kernels whose properties are demonstrated using several examples. In warping, a suitable transformation function is applied to map the input space into a new space where a standard kernel, e.g. Mat\'ern family of kernels, is able to predict the discontinuous function well. Our experiments show that these techniques have superior performance to GPs with standard kernels in modelling step-discontinuities.
\section*{Acknowledgements}
The authors gratefully acknowledge the financial support of the EPSRC via grant EP/N014391/1.
	
\bibliography{biblio}
\bibliographystyle{plain}
 \begin{appendices}
\section{Covariance functions/kernels} 
\label{sec_kernel}
Covariance kernels are positive definite (PD) functions. The symmetric function $k : \mathcal{D} \times \mathcal{D} \mapsto \mathbb{R}$ is PD if 
\begin{equation*} 
\sum_{i = 1} ^N	\sum_{j = 1} ^N \alpha_i \alpha_j k(\bx^i, \bx^j) \geq 0
\end{equation*}
for any $N \in \mathbb{N}$ points $\bx^1, \dots, \bx^N \in \mathcal{D}$ and $\boldsymbol{\alpha} = [\alpha_1, \dots, \alpha_N ]^\top \in \mathbb{R}^N$. If $k$ is a PD function, then the $N\times N$ matrix $\mathbf{K}$ whose elements are $\mathbf{K}_{i j} = k(\bx^i, \bx^j)$ is a positive semidefinite matrix because $\sum_{i = 1} ^N	\sum_{j = 1} ^N \alpha_i \alpha_j  \mathbf{K}_{i j} \geq 0$.

Checking the positive definiteness of a function is not easy. One can combine the existing kernels to make a new one. For example, if $k_1$ and $k_2$ are two kernels, the function $k$ obtained by the following operations is a valid covariance kernel: 
\begin{align*}
&k(\bx, \bx^\prime) = k_1(\bx, \bx^\prime) + k_2(\bx, \bx^\prime) \\
&k(\bx, \bx^\prime) = k_1 (\bx, \bx^\prime) \times k_2(\bx, \bx^\prime) \\
&k(\bx, \bx^\prime) = c k_1(\bx, \bx^\prime) ,~  c \in \mathbb{R}^+ \\
&k(\bx, \bx^\prime) = k_1(\bx, \bx^\prime) + c ,~  c \in \mathbb{R}^+ \\
&k(\bx, \bx^\prime) = g(\bx) k_1(\bx, \bx^\prime) g(\bx^\prime)~ \text{for any function} ~ g(.) .
\end{align*}
We refer the reader to \cite{GPML, duvenaud2014} for a detailed discussion about the composition of covariance functions. It is also possible to compose kernels with a function as explained in Section \ref{sec_warping}.

Usually, a covariance function depends on some parameters $\mathbf{p}$ which are unknown and need to be estimated from data. In practice, a parametric family of $k$ is chosen first. Then the parameters are estimated via maximum likelihood (ML), cross-validation or (full) Bayesian approaches \cite{GPML}. In the sequel, we describe the ML approach as is used in this paper. 

The likelihood function measures the adequacy between a probability distribution and the data;  a higher likelihood function means that observations are more consistent with the assumed distribution. In the GP framework, as observations are presumed to have the normal distribution, the likelihood function is
\begin{equation}
p \left( \mathbf{y} \vert \mathbf{X}, \mathbf{p}, \mu \right) = \frac{1}{(2\pi)^{n/2} \vert \mathbf{K}\vert ^{1/2}}\exp \left(- \frac{\left(\mathbf{y} - \mu \mathbf{1} \right)^\top \mathbf{K}^{-1} \left(\mathbf{y} - \mu \mathbf{1} \right)}{2} \right) ,
\end{equation}
where $\vert \mathbf{K}\vert$ is the determinant of the covariance matrix. In the above equation, if $\mu$ is unknown, it is replaced with its estimate given by Equation (\ref{mu_estim}). 

Usually for optimisation, it is more convenient to work with the natural logarithm of the likelihood (log-likelihood) function which is
\begin{equation}
\ln p \left( \mathbf{y} \vert \mathbf{X}, \mathbf{p}, \mu \right) = -\frac{n}{2} \ln(2\pi) - \frac{1}{2} \ln\vert \mathbf{K}\vert  - \frac{\left( \mathbf{y} - \mu \mathbf{1} \right)^\top \mathbf{K}^{-1} \left(\mathbf{y} - \mu \mathbf{1} \right)}{2} .
\label{log_lik}
\end{equation}
Maximising (\ref{log_lik}) is a challenging task as the log-likelihood function is often nonconvex with multiple maxima. To do so,  numerical optimisation algorithms are often applied.  We refer the reader to \cite{lophaven2002, forrester2008} for further information.
\end{appendices}
\end{document}